\documentclass[runningheads]{llncs}
\usepackage{graphicx}
\usepackage{todonotes}
\usepackage{hyperref}
\usepackage{cleveref}
\usepackage{subcaption}
\usepackage{microtype}
\usepackage{color,soul}
\usepackage{cite}
\begin{document}
\title{Voter Perceptions of Trust in Risk-Limiting Audits}
%
%
%
\author{Asmita Dalela, Oksana Kulyk, Carsten Sch{\"u}rmann}
%
\institute{IT University of Copenhagen, Denmark \\
\email{\{asmd,okku,carsten\}@itu.dk}}
%
\maketitle              
\begin{abstract}
  Risk-limiting audits (RLAs) are expected to strengthen the public confidence in the correctness of an election outcome.  We hypothesize that this is not always the case, in part because for large margins between the winner and the runner-up, the number of ballots to be drawn can be so small that voters lose confidence. We conduct a user study with 105 participants resident in the US. Our findings confirm the hypothesis, showing that our study participants felt less confident when they were told the number of ballots audited for RLAs. We elaborate on our findings and propose recommendations for future use of RLAs. 
  

\end{abstract}

\section{Introduction}


A credible election does not only produce an election result, it also provides guarantees that this election result is correct. Depending on the country and the electoral, these guarantees can take many different forms. In some countries, votes are counted more than once, for example in Scandinavia. In others, votes are counted only once and guarantees are given by a post-election audit, triggered either automatically, or when the margin between winner and runner-up is thin. One particular post-election audit, a so-called Risk-Limiting Audit (RLA), is gaining increasing popularity, because it can be extremely efficient especially if the margins are wide. It will automatically trigger a full recount if enough statistical evidence is not found to support the election outcome, and it can be adopted to several electoral systems, including first-past the post~\cite{lindemanStark12}, d'Hondt based systems~\cite{stark15jets}, and instant runoff elections~\cite{Blom19}. One would expect RLAs to strengthen public confidence in the election, first, because they can be integrated into existing election processes, for example, in Denmark, where the result of the first (rough) count can be verified during the second (fine) count~\cite{Schuermann16evoteid}. Second, some of the ceremonies surrounding RLAs can be turned into public events, such as the dice-rolling ceremony used to create entropy to select a random sample (see the public notice of the Colorado Secretary of State~\cite{Griswold20}). 

But do RLAs really strengthen public confidence? Of course, from a statistical point of view, they do. The theory is sound and the sample size is mathematically determined once the (diluted) margin and the risk limit are known. And yet, often, the sample size is ridiculously small, especially when the margins are wide, leading us to wonder if voters find those numbers convincing and confidence-raising.  In Denver County, Colorado, USA, where nearly 393,826 votes were cast during the 2020 election, the RLA that was conducted shortly after the election required a sample size of 523 votes to achieve a 96\% confidence (risk limit 0.04) that the election result was correct. In Kenya, during the Presidential Election of October 2017, where 15,593,050 votes were cast, a hypothetical RLA would 
need  to sample  only 166 ballots to achieve a confidence of 99\% (risk limit 0.01) in a best case scenario.  To answer the research question \emph{if RLAs really strengthen public confidence in the outcome of an election}, we conducted a user study with 105 randomly chosen US residents across all demographics using the Prolific platform. The participants in this study confirm several hypothesis, such as $H_{1,1}$ where we asked the participants about their opinion on the number of ballots to be selected for auditing, they provided a number higher than the number prescribed by the RLA methodology and $H_{1,2}$ where we found that the participants’ confidence in the audit results changed when they were informed about the number of ballots selected for auditing. The quantitative and qualitative analysis confirmed both hypotheses.

In this paper we describe the user study and its results.  We give a brief introduction to RLAs and how they work in Section~\ref{sec:background}, before we outline the methodology for this user study in Section~\ref{sec:methodology}. We report on results in Section~\ref{sec:results} and assess the impact of our findings in Section~\ref{sec:discussion}.


\section{Background}
\label{sec:background}

Post-election audits are a common part of election cycles around the world. They are designed to inform officials if there has been a problem with electronic voting or counting machines, they can act as a deterrent against fraud, and overall they are expected to increase public confidence into the election result. 

Trust and public confidence in correct outcomes, a  human trait that is particularly affected by the use of (election) technologies whose inner workings are neither transparent nor easily understandable, have been studied in general terms~\cite{Simon13} and in terms of Internet Voting~\cite{Markussen14evote,Schuermann16ipsa,spycher2011transparency,solvak2020does}, where there are no paper ballots available.  If paper ballots (hand-marked or machine-marked) are available,  they can be audited in different post-election audits, which we discuss next.

What constitutes a post election audit differs from country to country. In some countries, e.g., in Denmark, Germany, Norway,  post-election audits are an integral part of the counting and tabulation  process, and executed every time an election is held. In other countries there are clear rules when a recount is triggered, (a) if the margin is below 0.5\%, (b) at the discretion of the Secretary of State,  or (c) if voters file a petition. In the US, each state has their own local regulatory framework that defines who decides if a recount takes place, how many ballots are to be audited, from where  these ballots are drawn, and who is going to pay for the recount.  Lastly, there are many countries in the world, especially in developing and post-conflict countries, where there are no provisions for recounts at all.

In 2012, Lindeman and Stark~\cite{lindemanStark12}, devised a novel statistic-based method for election auditing, which is called a risk-limiting audit (RLA).  The advantages of an RLA are plentiful. It is the only method that can provide statistically valid evidence supporting the correctness of an election outcome, and it is also the only method that can correct an erroneous election outcome.  Without going into details or the different flavors of RLAs that exist, an RLA requires voter verified paper ballots, i.e.\ ballots that represent the intent of the voter, and it requires that the integrity of the paper trail is established and trusted.

An RLA is administered by the following procedure:  Given a risk-limit that defines the likelihood with which the RLA will recognize and correct an erroneous election outcome, and given the smallest margin between winners and losers, the RLA  then (1) computes  the sample size of ballots to be drawn \emph{random}, where it has become customary to create the entropy  using several 10-sided dice, and (2) identifies the individual ballots to be drawn.  RLAs can be used when ballots are identifiable, for example in UK elections where ballot papers are numbered or when ballots are sorted into batches. The actual audit consists of locating the physical ballots in the random sample, and then checking if they are correctly interpreted (digitally) or correctly sorted into batches.

RLAs were introduced in 2012 and are now regularly used in different US states for auditing US presidential, congressional, and local elections. One of the reasons why RLAs are used predominately in the US, is because ballots usually contain multiple races, which renders manual counting and tabulation impossible. Since ballots are digitally interpreted and electronically counted, RLAs provide a transparent way to ensure that the results are correct. Recent election observation reports~\cite{OSCEUSA18,OSCEUSA20} noted a trend towards hand-marked paper ballots and RLAs. 
According to the National Conference of State Legislatures~\cite{NCSL21}, the use of RLAs is in statute for the three states, Colorado, Rhode Island, and Virginia. 
The statutory pilot programs are run in Georgia, Indiana, and Nevada 
whereas the use of RLAs is optional in California, Ohio, Oregon and Washington. In contrast, Michigan and New Jersey are running only administrative pilot programs, similar to the one, we conducted in Denmark~\cite{Schuermann16evoteid}. For a detailed description of risk-limiting audits and its application in other countries, see~\cite{ifesrla21}.

Risk-limiting audits are currently considered as the gold standard among post-election audits because it is 
statistically sound, verifiable, and reproducible, given the entropy used for selecting the random sample.  However, besides all the mathematical rigor, there is also a psychological side to an RLA: If the respective margin between winners and losers 
 is wide, the number of ballots to be audited is actually really small.
 For example, in Kenya, the Presidential Election of 2017 reported that Uhuru Kenyatta obtained 8,223,369 votes, whereas his opponent Raila Odinga obtained 6,822,812 votes, which yields a margin of 1,400,557 votes.  When setting the risk-limit to 1\%, the ballot-level comparison RLA reports a sample size of 166 ballots. For some, considering that 15,593,050 ballots were cast, drawing only 166 ballots to verify such an important election may sound unbelievable, untrustworthy, and perhaps even unacceptable.



\section{Methodology}
\label{sec:methodology}
We describe the online survey, including the study hypotheses. 

\subsection{Study procedure}

The study was conducted as an online survey\footnote{We used the SoSciSurvey platform (\url{https://soscisurvey.de}) for hosting the survey} and consisted of the following parts:

\subsubsection{General audits} The participants were presented with a scenario about a hypothetical election 
with the following description:

\begin{quote}
\emph{Consider the following scenario: an election was conducted, with a total of [\texttt{NUM\_VOTERS}] voters casting their votes in favour 
of either candidate A or candidate B. After the count, candidate A got [\texttt{MARGIN}] votes more than candidate B, so candidate A was announced as 
the winner of the election.}
\end{quote}

The values of \texttt{NUM\_VOTERS} and \texttt{MARGIN} were dynamically generated for each new participant at random, with \texttt{NUM\_VOTERS} uniformly generated from the range between $2,900,000$ and $3,100,000$, and \texttt{MARGIN} uniformly selected to be in the range 
of 0\% to 20\% of the total vote (that is, modeling elections with two candidates where the share of votes for a winning candidate is between 50\% and 60\% of total cast votes)\footnote{The purpose of dynamically setting these values was to investigate the participants' answers based on a variety of margins.}.After presenting the scenario, the participants were asked whether they believed that an election audit would be a good idea if the candidate they supported lost the described election (Likert 5-point scale, from ``definitely not'' to ``definitely yes'') and were asked to explain their answer.

\subsubsection{RLA}

Afterwards, the participants were presented with the same election scenario once again. This time, they were being told that the election officials are planning to conduct a risk-limiting audit:

\begin{quote}
    \emph{After the election, the (election) authorities decided to conduct an audit known as a Risk Limiting Audit (RLA). This audit manually reviews a sample of ballots, to check whether the reported election result is correct. The number of ballots for this sample is not fixed but depends on the difference between the votes of the two candidates.}
\end{quote}

The participants were then asked whether such an audit would strengthen their confidence in the election result (Likert 5-point scale, from ``definitely not'' to ``definitely yes''), and how 
much time they believed it should take for the election officials to announce the results of the audit. Afterwards, they were asked to estimate the minimum number of ballots that should be sampled for such an audit, providing a number from $0$ to \texttt{NUM\_VOTERS}. Following that question, the participants were provided with the following description of the audit procedure:

\begin{quote}
    \emph{The auditors draw and inspect a random sample of [\texttt{NUM\_AUDITED}] ballots and conclude that they are 99\% certain that the election outcome is correct.}
\end{quote}

The value of \texttt{NUM\_AUDITED} was dynamically computed based on the values of \texttt{NUM\_VOTERS} and \texttt{MARGIN}, following the methodology described in \Cref{sec:background} to ensure the 99\% confidence, which corresponds to a risk limit of $1\%$.
The participants were asked once again, whether such an audit would improve their confidence in the election result (Likert 5-point scale, from ``definitely not'' to ``definitely yes'') and were requested to explain their answer.

\subsubsection{Selection criteria}

In the next part, the participants were presented with a list of criteria that could have been used for choosing the number of audited ballots, namely, (a) \emph{recommendation by NGOs and international organizations}, (b) \emph{existing legislation}, (c) \emph{methodology described in a scientific paper, openly available online}, (d) \emph{court decision}, (e) \emph{mutual agreement among all the political parties involved in the election} and (f) \emph{recommendation by independent experts}. For each of the criteria, the participants were asked how their reliance on it would affect their confidence in the election results 
(Likert 7-point scale, from ``I would be much less confident'' to ``I would be much more confident'') and were requested to explain their answer. The participants were then asked to provide other criteria they could think of that might be important to them as free-text answers.\footnote{For the sake of brevity, we omit the analysis of this part and provide it in the extended version of our paper.}

\subsubsection{Demographics} To conclude the survey, the participants were asked about their demographics, namely, gender, age, education, country of residence, and, if they are registered as a voter in the US elections, in which state and for which party they are registered. They were also asked whether they had any further remarks regarding the survey, and to elaborate as a free-text answer.

\subsection{Research questions}
\label{sec:hypotheses}
The aim of our study was to investigate the voters' mental models of election audits, in particular, focusing on the misalignment of the assurances provided by the RLAs and their perceptions among the potential voters. Specifically, we investigate whether the number of ballots required to be chosen by the RLA methodology is considered satisfactory by the voters, formulating the following main hypotheses for the \emph{quantitative evaluation}:

\begin{description}
\item[$\mathbf{H_{1,1}}$] When asked about their opinions about which number of ballots should be selected for auditing, the participants provide a number higher than the one prescribed by the RLA methodology.
\item[$\mathbf{H_{1,2}}$] Participants' confidence in the audit results changes when they are informed about the number of ballots selected for auditing

\end{description}

In addition to these hypotheses, we consider further aspects that might affect the voters' confidence in the election audits. We consider the effects on voters' confidence in the election audits in cases when the methodology of the audits is supported by one or more of the \emph{selection criteria}, that is, entities or processes presented to the participants in the surveys as criteria for choosing the number of ballots that should be audited. Finally, given the political landscape of the US elections, we study whether the party affiliation has an effect on the acceptance of audits. Thereby resulting in the following additional hypotheses:

\begin{description}
\item[$\mathbf{H_{2}}$] There is a difference in the effects on the voters' confidence in audits, depending on which selection criteria backs up the chosen number of audited ballots.
\item[$\mathbf{H_{3,1}}$] Party affiliation has an effect on whether the voter believes that election audits in general would be a good idea
\item[$\mathbf{H_{3,2}}$] Party affiliation has an effect on whether the voter believes that conducting RLA would strengthen their confidence in the election result \emph{before} being told the number of audited ballots
\item[$\mathbf{H_{3,3}}$] Party affiliation has an effect on whether the voter believes that conducting RLA would strengthen their confidence in the election result \emph{after} being told the number of audited ballots
\end{description}



In addition to this, we do a qualitative evaluation of open-ended answers to understand the attitudes towards election audits. 

\subsection{Recruitment and ethical considerations}

The survey was conducted in March 2021, and the participants for the study were recruited using the Prolific platform\footnote{\url{https://prolific.co}}. The recruitment was done in two stages: (1) in the \emph{pilot stage}, five participants were recruited to verify that there were no significant issues with the survey, and (2) in the \emph{full study} stage, 100 additional participants were recruited. As the pilot stage did not reveal any issues, the survey was not changed and the five participants from that stage were included in the overall evaluation. As we wanted to study the attitudes in the context of the US election, only participants with USA as their country of residence were allowed to participate in the survey. The survey was estimated to take 10 minutes, and participants were paid 1.25 GBP for their participation, which corresponds to the recommended reimbursement by the Prolific platform.

Prior to starting the survey, the participants were presented with a consent form, outlining the goals of the survey, stating that the survey will not collect any personally identifiable data and the anonymized results will be published in scientific reports. The participants were furthermore told that they can withdraw from the survey at any time without explanation, however, in that case they would not be able to get the reimbursement for their participation.


\section{Results}
\label{sec:results}

A total of 105 participants completed the survey, of them 46 women, 57 men, one non-binary person, and one participant who preferred not to report their gender. The majority of the participants reported having either a Bachelor (44 participants) or a Master (22 participants) degree. The most commonly represented age group was 30-34 years (24 respondents), followed by 25-29 (20 participants) and 35-39 (16 participants). Almost a half (50 participants) reported being registered Democrats, 23 were registered as Republican and 21 as Independent, the rest of the participants chose not to report their party affiliation.

\subsection{Quantitative evaluation}

We report on the results of evaluating the hypotheses defined in Section 3.2, as well as provide other descriptive statistics from the study. The p-values for all the hypotheses $H_{1,1}$, $H_{1,2}$, $H_{2}$, $H_{3,1}$, $H_{3,2}$, $H_{3,3}$, $H_{3,4}$ are adjusted for multiple comparison using the Bonferonni-Holm method. We report the adjusted values. Note, that we report on some statistical tests not included in our hypotheses outlined in Section 3.2, but nonetheless performed to get a better understanding of our sample; these are not included in the p-value adjustment. All the statistical analysis computations are performed using R packages ``rstatix'', ``coin'' and ``PMCMR''. We did not exclude any participants from the evaluation of the hypothesis $H_{1,2}$ and $H_{2}$, and excluded participants who did not provide a meaningful answer to the question that was critical for the evaluation from the analysis of the hypothesis $H_{1,1}$, $H_{3,1}$, $H_{3,2}$, $H_{3,3}$ which we describe in more details below.

\subsubsection{Number of audited ballots}

\paragraph{Preferred vs. actual number of audited ballots} For the comparison between the preferred number of audited ballots and the actual number (derived according to the RLA methodology) presented to the participants after they expressed their preference, we only considered participants who, when asked about the number of ballots that they believed should be audited, reported a number larger than 0 (thus excluding participants who either skipped the question or possibly believed that the RLA should not be conducted at all). After excluding 18 participants who did not answer that question and 4 additional participants who input 0 as their answer, 83 participants were included in the evaluation.

\begin{figure}
    \centering
    \includegraphics[width=0.68\textwidth]{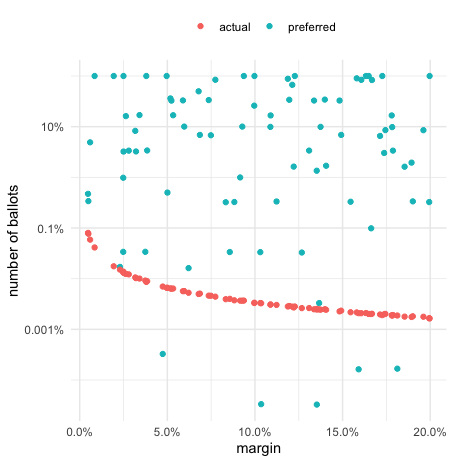}
    \caption{Preferred (as answered by participants) and actual audited ballots (as percentage of total ballots) depending on the margin. The scale is logarithmic. \label{fig:rla_ballot_difference}}
\end{figure}

When asked which number of ballots the participants would prefer to get audited, this number tended to be magnitudes higher than the actual number required by the RLAs for most of the participants (see \Cref{fig:rla_ballot_difference}). While the actual audited number of ballots presented to the participants ranged from .0016\% to .08\% out of total cast ballots (that is, from 51 to 2372 ballots respectively)\footnote{Note that the participants had similar number of total ballots presented to them, namely, ranging from 2.9m to 3.1m, and with the margin between 0 and 20\% of the total number of ballots.}, only 6\% of participants (5 out of 83) reported preferring a number of ballot less than or equal to the actual number required by the RLA methodology. Overall, the median preferred number of ballots was 6.9\%, and the standard deviation was 36.3\% of total number of ballots, indicating not only that participants overestimated the number of ballots required for audits, but also that they had highly varying opinions on what this number should be. The majority of the participants (57\%, 48 out of 83) preferred to audit less than 10\% of the total cast ballots, while 21\% (18 out of 83) preferred for more than half of the total cast ballots to be audited, and of them, the majority (12 participants, 14\% of 83) reported preferring a full recount.

The sign test\footnote{The test was chosen due to non-symmetrical distribution of the underlying data.} 
has confirmed that the difference between preferred and actual number of ballots is significantly different from zero ($p < .001$, 95\% CI for median difference between preferred and actual ballots (as percentage of total ballots) is [3.23\%, 16.1\%]), thus, $\mathbf{H_{1,1}}$ \textbf{is confirmed}.

\begin{figure}
    \centering
    \includegraphics[width=0.68\textwidth]{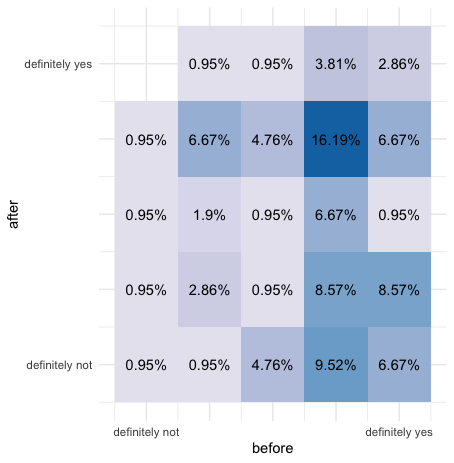}
    \caption{Percentage of participants choosing each of the combinations of their answers on whether RLAs would strengthen their confidence in the election result \emph{before} and \emph{after} seeing the number of audited ballots.\label{fig:rla_confidence_change}}

\end{figure}

\paragraph{Changes in confidence in RLAs} While the majority of the participants (70\%, 74 out of 105) had a positive attitude towards conducting RLAs, choosing either ``maybe yes'' or ``definitely yes'' as the answer to the question whether their confidence in the election result would increase based on the audits, only 44\% provided a positive answer to the same question asked after presenting the number of audited ballots to the participants. Consequently, while only 17\% of the participants (18 out of 105) provided a negative answer (either ``maybe no'' or ``definitely not'') before seeing the number of audited ballots, this percentage increased to 45\% (47 out of 105) after the participants were presented with that number. Overall, the majority of the participants (54\%, 57 out of 105), were  less likely to think that RLAs would increase their confidence in the election results when they were told the number of audited ballots (median decrease of 1 point on the 5-point scale). Furthermore, for 33\% of the participants (35 out of 105), being told the number of ballots led to changing their attitude towards RLAs from positive (answering either ``maybe yes'' or ``definitely yes'' when asked whether RLAs would increase their confidence in the election result) to negative (choosing the answers ``maybe no'' or ``definitely not''). \Cref{fig:rla_confidence_change} shows the distribution of changes of participants' answers. 

The Wilcoxon signed-rank test shows a significant difference between the ``before'' and ``after'' answers ($p < .001$, $Z = -4.47$, effect size $r = .33$, moderate), thus, $\mathbf{H_{1,2}}$ \textbf{is confirmed}. Furthermore, Spearman's test did not show significant correlation between confidence change (as the difference between the ``before'' and ``after'' answers) and either the number of audited ballots ($p = .1863$, $r = .13$) or the margin ($p = .185$, $r = -.13$).

\subsubsection{Selection criteria}

\begin{figure}[htb]
    \centering
    \includegraphics[width=0.7\textwidth]{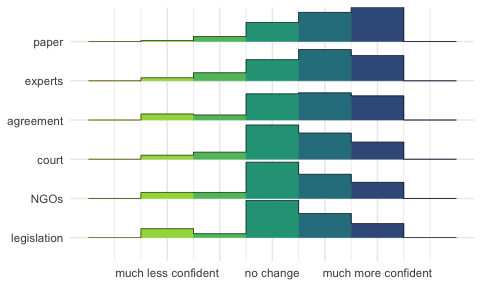}
    \caption{Answers to the question, how choosing the number of audited ballots based on the following selection criteria would affect the participants' confidence in the election result.}
    \label{fig:rla_trust_sources}
\end{figure}

For all of the selection criteria for the number of audited ballots outlined in \Cref{sec:methodology}, namely, {recommendation by NGOs and international organizations} (``NGO''), {existing legislation} (``legislation''), {methodology described in a scientific paper, openly available online} (``paper''), {court decision} (``court''), {mutual agreement among all the political parties involved in the election} (``agreement'') and {recommendation by independent experts} (``expert''), most of the participants reported that their confidence would either improve or stay the same. Namely, the median score for the selection criteria ``paper'',  ``expert'' and ``agreement'' was 4, meaning that the majority of the participants answered that they would feel either ``much more confident'' or ``more confident'' -- if the number of audited ballots was chosen according to these criteria. For the rest of the selection criteria, the median value was 3, meaning that the majority of the participants selected one of the options ``much more confident'', ``confident'' or ``my confidence would not change''. At the same time, the number of participants that answered that they would feel either ``less confident'' or ``much less confident'' if the number of audited ballots was based on specific selection criteria ranged from 5.7\% (6 out of 105) for the methodology described in a scientific paper to 12\% (13 out of 105) for existing legislation. The summary of the participants' answers is provided on \Cref{fig:rla_trust_sources}.

The Friedman test comparing the scores for all the selection criteria resulted in $p < .001$, with $\chi^2 = 30.863$ and the effect size $W = .07$ (small). Thus, $\mathbf{H_{2}}$ \textbf{is confirmed}. The post-hoc tests\footnote{The results of post-hoc tests here and in the next subsection are not included in the p-value adjustment} show significant differences between ``paper'' and ``legislation'' ($p = .0017$), ``paper'' and ``NGO'' ($p = .009$) and ``paper'' and ``court'' ($p = .03$).



\subsubsection{Party differences} For the evaluation of the effects of political views on attitudes towards audits and selection criteria, we exclude participants who chose not to provide their party affiliation (11 participants), resulting in 94 participants for the evaluation.

\begin{figure}[t]
\includegraphics[width=1\textwidth]{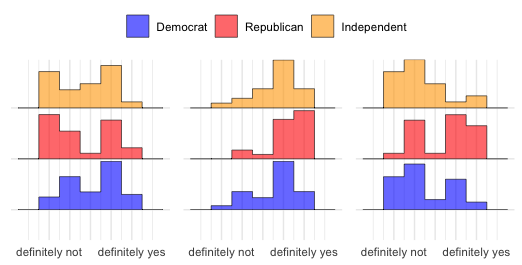}
\caption{Answers to the question (left to right), whether (a) participants believed that conducting audits would be a good idea if their candidate lost the election, (b) conducting risk-limiting audits would increase the participants' confidence in the election results \emph{before} specifying the number of ballots to be audited, (c) whether conducting risk-limiting audits would increase the participants' confidence in the election results (\emph{after} specifying the number of ballots to be audited.}%
\label{fig:rla_party_differences}
\end{figure}

\paragraph{Attitudes towards auditing in general} While only 34\% of all the participants who specified their party affiliation (32 out of 94) had positive views on audits in case their candidate lost the election (answering either ``definitely yes'' or ``mostly yes'' to the question whether such an audit would be a good idea), this number was higher among the participants who reported Republican as their party affiliation (61\%, 14 out of 23) than among Democrats (30\%, 15 out of 50) and Independents (14\%, 3 out of 21), see also \Cref{fig:rla_party_differences}. Similarly, only 35\% of Republicans (8 out of 23) provided a negative answer to the same question (either ``definitely not'' or ``mostly not''), compared to 62\% of Democrats (31 out of 50) and 67\% of Independents (14 out of 21). The Kruskal-Wallis test confirms the significant difference between the groups ($p = .015$, $\chi^2 = 10.579$, effect size $\eta^2 = 0.9$, moderate), thus \textbf{confirming} $\mathbf{H_{3,1}}$. The post-hoc tests furthermore show significant differences between Republicans and both Democrats ($p = .014$) and Independents ($p = .016)$.

\paragraph{Attitudes towards RLAs before specifying the number of audited ballots}

Most of the respondents across all parties had a positive attitude towards conducting RLAs in general, choosing the option either ``maybe yes'' or ``definitely yes'' when asked whether conducting such audits would strengthen their confidence in the election result (66\% of Democrats, 33 out of 50; 87\% of Republicans, 20 out of 23; 66.7\% of Independents, 14 out of 21), with participants who reported Republican as their party affiliation being the most likely to provide positive results (see also \Cref{fig:rla_party_differences}). At the same time, only 9\% of Republicans (2 out of 23) and 14\% of Independents (3 out of 21) expressed negative attitudes towards audits in such a scenario, answering either ``maybe no'' or ``definitely not'', compared to 22\% of Democrats (11 out of 50). The Kruskal-Wallis test furthermore shows significant difference between the groups ($p = .04$, $\chi^2 = 7.8192$, effect size $\eta^2 = .06$, moderate), \textbf{confirming} $\mathbf{H_{3,2}}$, and post-hoc tests showing significant difference between Republicans and Democrats ($p = .029)$.

\paragraph{Attitudes towards RLA after specifying the number of audited ballots} Less than half of the participants reported believing that their confidence in the election result would increase after conducting RLAs with the specified number of ballots, answering either ``definitely yes'' or ``maybe yes''; of them, 50\% of Democrats (25 out of 50), 39\% of Republicans (9 out of 23) and 36\% of Independents (8 out of 21), see also \Cref{fig:rla_party_differences}. Consequently, a large percentage of participants across all parties held a negative view, with 36\% of Democrats (18 out of 50), 56\% of Republicans (13 out of 23) and 42\% of Independents (9 out of 21) answering either ``maybe not'' or ``definitely not''. Nonetheless, the Kruskal-Wallis tests did not reveal any significant difference between the groups ($p = .15$, $\chi^2 = 3.6773$), thus \textbf{failing to confirm} $\mathbf{H_{3,3}}$.

\subsection{Qualitative evaluation}

We report the results of analyzing the open-ended answers of the survey. The answers were analyzed by two paper authors using open coding, and the codes were iteratively discussed until agreement was reached. We report on our findings in the subsections below. Each of the reported codes was mentioned at least by two participants. Since the goal of the qualitative evaluation is to understand the range of possible opinions, we do not report on the number of participants mentioning each code, instead using quantifiers according to \Cref{tbl:quantifiers}. 

\begin{table}[htb]
\centering
\begin{tabular}{rr}
  \hline
$< 20\%$ & few \\
$< 40\%$ & some \\
$< 60\%$ & many \\
$< 80\%$ & most \\
$\geq 80\%$ & near-all \\
   \hline
\end{tabular}
\caption{Quantifiers used in describing the qualitative results, by percentage of participants mentioning each code.\label{tbl:quantifiers}}\end{table}

\subsubsection{Attitudes towards auditing} We describe the codes identified when studying the open-ended answers of participants regarding their attitudes towards auditing. Namely, we consider their opinions regarding (1) whether audits at all would be a good option, (2) whether risk-limiting audits in particular would increase their confidence, \emph{without specifying the number of audited ballots}, and whether risk-limiting audits \emph{with a specified number of audited ballots }would increase their confidence. For each of the three questions, we group the participants into the ones who expressed a \emph{negative} (selecting either option 1 or 2 on the corresponding 5-point Likert scale), \emph{neutral} (selecting option 3) or \emph{positive} (selecting either option 4 or 5) opinion.

\paragraph{\textbf{Audits in general}}

Among the participants who expressed a \textbf{negative} opinion (58 participants) towards audits, most mentioned the fact that the \emph{margin was sufficiently wide}. Few  participants mentioned that given such a margin an \emph{audit is unlikely to change anything}, or that \emph{fraud or errors of that scale would be very unlikely}. Few participants mentioned \emph{negative effects of audits}, such as delays in announcing election results, monetary costs and overall loss of confidence. Few of the participants mentioned that they \emph{trust the election system}, or that an audit is not necessary unless \emph{suspicious activity takes place}.

Considering the answers from the participants with a \textbf{neutral} opinion (11 participants) towards audits, some mentioned the fact that the \emph{audit is needed if there is inconsistency or fraud in the election process}. Few participants liked the idea of audits but doubted the audit process, by articulating that \emph{not enough information is given to decide need for audit}. Few  participants expressed that audits will not be useful as they feel confident that \emph{election was free and fair} and need more information to decide the need for an audit.

The participants with \textbf{positive} opinion (36 participants) towards audit, many remarked that \emph{audit is good for reconfirming election results} regardless of which party wins the election. Few  participants also emphasized the need of audit by stating that it will help in checking the errors which occur during the elections such as \emph{flaws in the way the votes are counted} and it can provide a \emph{check against irregularities that stood out from past elections}. These participants opined that audits will increase the accuracy of the results. Also, a few participants mentioned that audits are necessary when the \emph{margin is too small} to bring confidence in the election results.

\paragraph{\textbf{Risk-limiting audits}}

Out of  18 participants with negative attitudes towards RLAs, some mentioned that they believed the \emph{methodology is flawed} (e.g. saying that recounting only a sample of ballots is not enough or expressing doubts that sampling would be done in a proper way to ensure representative results, because of  either lack of due diligence or malicious intent on behalf of auditing authorities). Some stated that the \emph{margin was wide enough} so that no RLAs would be necessary. Few answered that they believed \emph{RLAs would not change anything} in the election result, or mentioned \emph{negative effects of the audits} such as loss of time or confidence in elections.

Among the participants with a \textbf{neutral} opinion (13 participants) towards RLAs, some mentioned that they \emph{need more clarity on the statistical model} behind the number of ballots selected, indicating that since RLA only reviews a sample of ballots; a possibility of error still exists. Furthermore, some mentioned that they \emph{ trust the election results} and feel that \emph{audit is not necessary} as it will not solve other issues related to voting such as gerrymandering, which are \emph{shady and need fixing}. Few participants also opined that \emph{audit will reconfirm the election results} but emphasized that it is \emph{not necessary to conduct}.

For the participants with \textbf{positive} opinion (74 participants), most mentioned the fact that the RLAs will help in \emph{confirming the validity of the election results} and give a feeling that everything was done correctly. Few of them mentioned that such \emph{reassurance of preventing irregularities} will help in building the \emph{trust in the election process}. On the other hand, few who liked the idea of audit, \emph{wanted more clarity on the statistical model used to determine the random sample} and mentioned that sample size needs to be a true reflection of the electorate.

\paragraph{\textbf{Risk-limiting audits with specified number of audited ballots}}

Out of  47 participants who expressed a negative opinion towards RLAs when shown the audited number of ballots, near-all answered that they believed that the \emph{audited number was too small}, also comparing it to the margin or the total number of cast ballots, as well as providing  such further explanations as the sampled number being within margin of error. Few, furthermore, mentioned that they \emph{did not believe the claims of 99\% accuracy} given the audit, or that they believed that auditing such a small number \emph{does not show a serious effort} on behalf of election authorities. Few participants furthermore stated that they believed that \emph{sampling is not good enough} to confirm the election result (as opposed to full recount), or, on the contrary, that \emph{audits would not be necessary at all} in the described scenario.

Among the participants with a \textbf{neutral} opinion (12 participants) towards RLAs when shown a specified number of audited ballots, most mentioned that the \emph{audited number was too small} and felt that it wouldn't affect their perception about the validity of the election results in any major way. Few participants further questioned the need for conducting such audits by emphasizing that since the \emph{margin of certainty is 99\% it makes little sense to conduct such audits} and expressed that they would like to see the statistical method used for computing the audited number of ballots.

For the participants with \textbf{positive} opinion (46 participants), many mentioned that the \emph{audit will reconfirm the election with 99\% certainty} and will increase their trust in the election process. A few opined that it will be value-adding if the \emph{audits are inspected by a third party to avoid potential frauds or discrepancies}. Few participants who showed an inclination for audits articulated the \emph{need for sampling a large number of ballots} to convince majority of people. On the contrary, a few of them were \emph{satisfied with the the sample size} chosen for the audit and stated that the \emph{RLA is representing a fair sample of ballots to be audited}.

\section{Discussion and Conclusions}
\label{sec:discussion}

If we consider the quantitative and qualitative analysis in conjunction, a clearer picture emerges: despite the statistical soundness of the RLA method, human factors pose additional challenges to building public confidence, which is one of the objectives of RLAs.  In this section, we describe these challenges and offer methods to respond to them.

\emph{Sample sizes.} The qualitative part of our study confirmed the hypothesis that the voters' expectation of how many ballots to audit exceeds the sample sizes computed by the RLA algorithm. We could also confirm the hypothesis that when sharing information about the size of the sample with the voter, the sample-size affects the voter's perception of what is an appropriate number of ballots to audit. We conclude that small sample sizes can cause distrust. Two ways to respond to this challenge come to mind. First, by auditing more ballots than required one can better align the sample size more closely to the expectations of the voter, lowering the risk for creating distrust.  
Second, by structuring a national RLA into smaller RLAs to be executed on the jurisdiction or even precinct level, one implicitly increases the number of ballots to be audited. This is what happened for the 2020 US Presidential election, albeit not by choice but by law.

\emph{Voter education.} Our analysis also stresses the importance of voter education. Most voters are not familiar with the statistical theory behind RLAs and therefore it is not necessarily the case that an RLA is effective in strengthening public confidence in the election outcome. To raise the effectiveness of RLAs for trust-building, targeted efforts should be undertaken to educate the public on how to read and interpret the data that is the result of an RLA.\footnote{See for example \url{https://sos.ga.gov/index.php/elections/2020_general_election_risk-limiting_audit}}

\emph{Paper trail integrity.} RLAs work under the assumption that the integrity of the paper trail is intact, which means that the paper trail is properly secured between vote casting and auditing. A few participants stated unprompted, how important the integrity of the paper trail actually is. We conclude that using hand-marked or even machine-marked paper ballots alone are not sufficient to strengthen public confidence and trust in the election, but additional efforts need to be undertaken to explain to the voter why the paper trail has integrity.

\emph{External factors.} We found that there are external factors that could strengthen the confidence in the correct choice of the sample to be drawn, and these factors include academic research papers, experts, and consensus among the  competing parties. We leave this as food for thought for election management bodies, who consider implementing RLAs as part of the election process.  

In future work, we propose to conduct additional user studies to investigate in more depth, the effects of auditing votes on a jurisdictional or even precinct level on public confidence, and the effect full hand-counts have on public trust. Paradoxically, we speculate that voters will trust several smaller RLAs more then one big RLA on the national level. It would also be interesting to study the effect of choosing larger sample size than the RLA-determined one to strengthen public confidence in the election outcome.  We realize of course, that this more a political science than a technical research question.

\textbf{Limitations: } Due to our use of crowdsourcing for recruitment of the participants, our sample might be biased towards participants younger, better-educated and more likely to be active Internet users than the general population. While studies show that crowdfunding platforms can provide a sample representative of this demographic \cite{redmiles2019well}, future studies need to be performed to study the perception of RLAs among the rest of the population. 

We considered the ballot comparison method for RLAs, which is one of two methods commonly used in the US. Future studies will be needed to investigate whether the other method, which also results in auditing a larger sample of ballots, is more likely to create trust.

In our study, we did not ask the participants about their prior knowledge about RLAs or attempted to educate them about the RLA procedure. While such investigations would be valuable point of future studies (e.g. in designing effective voter education measures), our aim was to measure the confidence in RLAs and how it is affected by publishing the number of audited ballots among potential voters in the US as it is at this point of time, without extra interventions from our side.

In our study we aimed to variate the number of sampled ballots (as a function of total number of cast ballots and margin), in order to study the participants' attitudes given a variety of scenarios. Nonetheless, our variations still ended up in a relatively small interval of possible sample sizes. As such, very small margins ($ < 1\%$), that would result in a large amount of audited ballots, were seldom present and very large margins ($>20\%$) were not present at all. Future works will address this. 

Finally, we presented an abstract and hypothetical election scenario, whereby trust in election outcome is affected by an overall context. Our findings however do show that the usage of RLAs alone is not enough to improve trust and might actually lower it.





\section{Conclusion}
We are not against RLAs. We do argue that as a measure to create trust, they are not sufficient by themselves, and additional measures such as voter education need to be considered. While this study is the first one to investigate this issue, follow-up work is needed to better understand the factors influencing voter's trust and the effectiveness of various ways one can educate voters about RLAs or raise trust via other measures.
\bibliographystyle{plain}
\bibliography{demtech}

\end{document}